\documentstyle[preprint,revtex]{aps}
\begin{document}
\draft
\begin{title}
Exact calculation of spectral properties of a particle interacting\\
with a one dimensional fermionic system.
\end{title}
\author{H. Castella$^*$, X. Zotos}
\begin{instit}
Institut Romand de Recherche Num\'erique en Physique des Mat\'eriaux,
(IRRMA)\\
PHB-Ecublens, CH-1015 Lausanne, Switzerland\\
{\rm and$^*$}\\
D\'epartement de Physique de la Mati\`ere Condens\'ee\\
24, quai E. Ansermet, CH-1211 Gen\`eve, Switzerland\\
\end{instit}
\begin{abstract}
Using the Bethe ansatz analysis as was reformulated by Edwards, we calculate
the spectral properties of a particle interacting with a
bath of fermions in one dimension for the case of equal particle-fermion
masses. These are directly related to singularities apparent in optical
experiments in one dimensional systems.
The orthogonality catastrophe for the case of an infinite particle mass
survives in the limit of equal masses.
We find that the exponent $\beta$ of the quasiparticle
weight, $Z\simeq N^{-\beta}$ is different for the two cases,
and proportional to their respective phaseshifts at the Fermi surface;
we present a simple physical argument for this difference.
We also show that these exponents describe the low energy behavior of the
spectral function, for repulsive as well as attractive interaction.
\end{abstract}

\pacs{78.70.Dm,71.35+z,75.10.Lp}

\section{Introduction}
The motion of a particle interacting with a Fermi bath has been the subject
of many studies as it relates to a variety of physical phenomena~:
muon diffusion in metals \cite{kon} , stability of the fully ferromagnetic
state (the Nagaoka problem) \cite{sha} and particularly relevant to
optical experiments in degenerate semiconductors and metals \cite{mnd,cal},
where the hole created by optical excitation plays the role of the particle
interacting with conduction electrons.

In the late 60's the Fermi edge singularities in the X-ray absorption spectra
of metals were explained in terms of correlation effects between
conduction electrons and the photon created hole; Mahan first predicted the
singularities before Nozi\`eres
and DeDominicis calculated the exact spectrum; these pioneering
works gave rise to the so called MND problem \cite{mnd}. Relying
on perturbative arguments valid for a three dimensional system,
it was believed that such singularities would disappear
in valence band excitations where the mass of the hole is comparable
to the conduction electron mass, in contrast
to the case of core level excitations, where the mass of the localized hole
can be considered infinite \cite{don}. In lower dimensions the situation
is different \cite{kop} and in fact strong enhancement at
threshold was recently observed in photoluminescence on
quantum wires \cite{cal}. These experiments have raised further interest
in optical singularities in low dimensional systems.

This model was also studied in connection with ferromagnetism in single
band models and it has recently gained renewed attention in two dimensions
because of the possible relevance of the single band Hubbard model to
describe the high-Tc superconductors.
In this context it describes the dynamics of a single spin flip in a
ferromagnetic background and adresses
the question of the stability of the Nagaoka state or the limitations
of Gutzwiller type variational wave functions used for this problem
\cite{sha}.

In this work we study the spectral properties of the particle in
one dimension and for the special
case where the mass of the particle is the same as that of the fermions
in the bath. For this analysis we use the analogy to McGuire's
solution of the problem
of a single spin flip interacting with a ferromagnetic background in a
one dimensional continuous system, considering the particle as the reversed
spin \cite{mcg}. Actually McGuire's work is a precursor,
for the continuum model,
of Lieb and Wu's Bethe Ansatz (BA) exact solution of the Hubbard
model in one dimension \cite{lie}. McGuire calculated static
correlation functions as well as the effective mass of the reversed spin;
a lattice version of the latter was presented in Ref. \cite{zot}.
However, despite
the BA solution, an exact calculation of the dynamic quantities
has not been possible as it involves the calculation of matrix elements
between, difficult to handle, BA wavefunctions for the excited states.
In this work we use a new presentation of McGuire's solution, due
to Edwards \cite{edw}; the relative simplicity of the wave
functions in this case allows us to evaluate directly the spectral
weight $Z$
and the spectral function $A(k,\omega)$ in the small momentum $k$ regime.
Our results show that the orthogonality catastrophe of Anderson
occurs \cite{and}~: a quasiparticle description of the excitations
in terms of non interacting eigenstates is therefore no more
possible; the spectral function is  totally incoherent and has the same
sort of divergence as in the infinite mass case.

The occurence of the orthogonality catastrophe and the divergence of the
spectral weight were recently predicted within perturbation theory for
the one dimensional problem \cite{kop}. The edge singularities of
the absorption spectrum were also derived for a Tomonaga-Luttinger model
and it was claimed that the corresponding exponents would not depend
on the mass of the hole \cite{oga}. In our work however, we show
that the exponents of the spectral function differ in
the two extreme limits of infinite mass and equal masses. This
seems in contradiction with the latter results; we will
discuss the importance of backscattering, which was omitted
in the Tomonaga-Luttinger model, to correctly describe
the case of an infinite particle mass.

We note that, except for the very special
case of one hole in a half filled band with infinite repulsion
\cite{sor}, our work presents the first calculation of
dynamical correlation functions using directly
the BA wavefunctions. Our calculations agree with previous
results by Frahm and Korepin \cite{fra} thus giving support to
the assumption of conformal invariance of the model used.

In summary we show in this article that the orthogonality catastrophe
takes place for the case of equal particle-fermions masses and
that the same exponents characterize the divergence of the spectral
function at threshold in the small momentum regime. The exponents
however differ in the two cases of infinite mass and equal masses.
The article is organized as follows.
In section 2 we present the model and its exact solution following
mainly Edwards.
Our analysis is performed for the continuous model and we simply quote the
results for the lattice. In section 3, we derive
the orthogonality catastrophe for the repulsive interaction.
In section 4, we calculate the spectral function for the repulsive
and attractive case; in the latter we discuss the influence of the
bound state.

\section{Model}
The model Hamiltonian describes $N$ fermions of mass $m$ and one particle
of mass $m_h$ interacting via a delta function potential and moving on a line
of length $L$ with periodic boundary conditions~:

\begin{equation}
H=-\frac{1}{2m}\sum_{i=1}^N\frac{\partial ^2}{\partial x_i^2}-
\frac{1}{2m_h}\frac{\partial ^2}
{\partial x_0^2}+U\sum_{i=1}^N\delta(x_i-x_0)
\end{equation}

Throughout this article we use the convention $m=1$ and $m_h \rightarrow
m_h/m$. When $m_h \rightarrow \infty$ the problem reduces to a
single particle problem. The fermions evolve in the static potential
created by the hole. The response of the Fermi sea to the sudden
appearance of the disturbance was calculated by Nozi\`eres and
De Dominicis\cite{mnd}. For a finite mass however, this model gives rise to
a full many body problem.

We will present at first the problem quite generally for an arbitrary mass and
then focus on the solution for $m_h=1$ following mostly Edwards presentation.
It is convenient to express the problem in the reference frame
attached to the particle. The wave functions in the two
reference frames are related by a simple translation~:

\begin{equation}
\Psi(x_0,\ldots,x_N)=e^{iQx_0}f(x_1-x_0,\ldots,x_N-x_0)
\end{equation}

\noindent
Q is the total momentum. For our periodic boundary
conditions $Q=2\pi m/L$, m being an integer, and
$y_j=x_j-x_0\in[0,L]$. The Schr\"odinger equation for f is then~:

\begin{equation}
\left(\frac{1}{2m_h}\left(Q+i\sum_{j=1}^N\frac{\partial}
{\partial y_j}\right)^2  -
\frac{1}{2}\sum_{j=1}^N\frac
{\partial^2}{\partial y_j^2}+U\sum_{j=1}^N\delta(y_j)\right)~f
=E~f
\end{equation}

\noindent
It describes the motion of $N$ fermions in a static local potential and
interacting via their total momentum.
In order to satisfy the equation everywhere and the periodicity, we add the
boundary conditions~:

\begin{eqnarray}
f(y_1,\ldots,y_N)\left|_{y_i=0}^L\right.= && 0 \nonumber\\
-\frac{1}{\mu}\frac{\partial}{\partial y_i}f(y_1,\ldots,
y_N)\left|_{y_i=0}^L \right.=
&& 2U f(y_1,\ldots,y_i=0,\ldots,y_N)
\end{eqnarray}

\noindent
where $\mu=m_h/(1+m_h)$ is the relative mass.
Edwards proposed the
following solution in the case $m_h=1$~:

\begin{equation}
f(y_1,\ldots,y_N)=\frac{1}{\sqrt{N!L}}\det(\phi_j(y_l))
\end{equation}

\noindent
For this wave function, the $\phi_j$ are normalized and
the boundary conditions imply~: $\phi_j(0)=
\phi_j(L)$ and $\phi_j'(0)-\phi_j'(L)=U\phi_j(0)$.
This can be achieved if the functions $\phi_j$ have the
following form~:

\begin{equation}
\phi_j(y)=\sum_{t=0}^{N}a_j^te^{ik_ty}
\end{equation}

\noindent
where the momenta $k_j$ are all different and solutions of the BA type
of equations~:

\begin{eqnarray}
Lk_j=2\pi\left(n_j+1/2\right)-2 && \arctan(2(k_j-\Lambda)/U),~~~j=0,...,N
\nonumber\\
Q=\sum_{j=0}^N k_j~~ && {\rm and}~~E=\frac{1}{2}\sum_{j=0}^N k_j^2
\end{eqnarray}

\noindent
These are Lieb and Wu's equations \cite{lie} in our special spin sector
of a single spin flip from the totally ferromagnetic state and in
the continuum limit; for an even total number of particles $N+1$, the $n_j$
are integers. A similar set of equations was originally derived by McGuire.
The eigenstates are specified by the quantum numbers $n_j,~~j=0,\ldots,N$
and m; $\Lambda$ is a real number which ensures that the total momentum is
$Q=2\pi m/L$ for a given choice of $n_j$.

We would like to focus on the simplicity of this solution compared
to McGuire's original one. Indeed the latter was written in the
static reference frame and defined by pieces in the different
regions $x_j<x_0<x_{j'}$ corresponding to a given
ordering of the particles on the line; moreover
it was a superposition of $N+1$ determinants. In contrast
Edwards solution is defined on the compact domain $[0,L]^N$ and
it takes the form of a single determinant. However,
we point out that the plane waves appearing in (6) are not
the usual free states (as $k_t\ne 2\pi n/L$ for $U\ne 0$)
and are not periodic on $[0,L]$ unless $U=0$.
But the functions $\phi_j$ have to be periodic.

In order to construct the total wavefunction, we have to
first solve the BA equations for a given choice of quantum
numbers, form the linear combinations in (6) and then build up
the determinant. We have $N+1$ plane waves and $N$ $\phi_j$'s.
Is the total wavefunction uniquely defined by this procedure ?
In fact there is an additional constraint on the coefficients
$a_j^t$ coming from the periodicity requirement~:

\begin{equation}
\sum_{t=0}^Na_j^t\left(1-e^{ik_tL}\right)=0
\end{equation}

\noindent
This restricts the $\phi_j$'s to a $N$ dimensional subspace
where we can make an arbitrary choice of basis. The total
wavefunction being a determinant is thus unique.

We can find a physically convenient basis of functions~:

\begin{eqnarray}
\phi_j(y)= && A_j (e^{i(k_jy+\delta_j)}-\sin(\delta_j)g(y))
,~~~j=1,\ldots,N\nonumber\\
g(y)= && \frac{\sum_le^{i(k_ly+\delta_l)}}{\sum_l
\sin(\delta_l)}
\end{eqnarray}
\noindent
where the phaseshifts are given by the BA equations~:
\begin{equation}
Lk_j= 2\pi n_j-2\delta _j, ~~~
\delta _j= -\frac{\pi}{2}+\arctan\left(\frac{2k_j-\Lambda}{U}\right)
\end{equation}

\noindent
$g(y)$ is an almost localized function for large $L$ and $N$;
it corrects the
plane wave around the origin in order to achieve periodicity without
affecting its plane wave character almost everywhere.
We show in the appendix the following statements :

\begin{equation}
A_j \rightarrow \frac{1}{L}(1+O(1/L))
\end{equation}
\begin{equation}
\int_0^L\phi_j^*(y)\phi_l(y)dy=\delta_{jl}+O(1/L)
\end{equation}
\begin{equation}
X_j^p= \frac{1}{\sqrt{L}}~\int_0^L e^{-i\frac{2\pi p}{L}y}\phi_j(y)dy=
\frac{\sin(\delta_j)}{(n_j-p)-\delta_j/\pi} + O(\log(L)/L),~~~p~{\rm
{}~integer}
\end{equation}

\noindent
These last results tell us that the $\phi_j$ behave like an orthonormal
set of single particle scattering states for large systems~:

\begin{equation}
\phi_j(y) \simeq \frac{1}{\sqrt{L}}e^{i(k_jy+\delta_j)}~,~~0\ll y\ll L
\end{equation}

\noindent
These are central results in our paper and the overlaps $X_j^p$
are used in all the subsequent calculations.

\section{Orthogonality catastrophe}

Anderson studied the influence of a static potential on a Fermi sea
\cite{and};
he showed that the ground state, $|\psi_0\rangle$, of the system
in the presence of the potential became orthogonal in the
thermodynamic limit to the ground state without potential,
$|\tilde \psi_0\rangle$~:

\begin{equation}
Z~ =|\langle\tilde \psi_0|\psi_0\rangle|^2
\propto N^{-\beta^+-\beta^-},~~~~
\beta^{\pm}=\left(\frac{\delta_F^{\pm}}{\pi}\right)^2
\end{equation}

\noindent
where $\delta_F^{\pm}$ are the phaseshifts of the electrons at the Fermi
surface for the two channels of even and odd parity wave functions, $N$
the number of fermions and the overlap $Z$ the spectral weight.
This orthogonality catastrophe takes place in the MND problem where the
static potential is created by the core hole. Moreover the same
exponents $\beta^{\pm}$ appear in the spectral function of the hole.
In our model with $m_h \rightarrow \infty$, the odd states do not
feel a potential located at the origin and the odd phaseshift vanishes;
only the even parity states
contribute to Anderson's orthogonality catastrophe.

In this section we calculate the spectral weight for our
model with equal masses and a repulsive interaction. We
show that the orthogonality catastrophe also takes place but with
different exponents than for an infinite mass.The spectral weight
can be evaluated with the use of our wave function
for the ground state and the corresponding overlaps $X_l^m$~:

\begin{equation}
Z= \left(\int dx_0\ldots dx_N\tilde f^*~f\right)^2
=\det(X_l^m)^2
\end{equation}

\noindent
$f$ is here the interacting ground state wavefunction in the
reference frame of the particle and $\tilde f$ the
corresponding wavefunction for the non interacting system.
The distribution of momenta for the ground state is given by the
following choice of quantum numbers~: the $n_j$ are
consecutive integers
from $-(N+1)/2$ to $+(N-1)/2$ and $\Lambda =0$; the
total momentum vanishes. Both $k$ and $-k$ appear in the spectrum
and it is natural to form even and odd combinations  of the
interacting single particle wavefunctions $\phi _j$; they are then
combinations of $\cos\left(k_jy+\delta_j\right)$ and
$\sin\left(k_jy+\delta_j\right)$ for the positive $k_j$.
We note here that both the even and odd channels have a non vanishing
phaseshift in contrast to the
infinite mass case. The non interacting states can also be expressed
as odd and even functions. The spectral weight factorizes then
into two contributions coming from both parities, $Z^+$ and
$Z^-$~:

\begin{equation}
\sqrt{Z^{\pm}}\simeq\left(\Pi_l
\frac{\sin\delta_l}{\pi}\right)\det\left(\frac{1}{(n-n')-\delta_n/\pi}
\right)
\end{equation}

\noindent
The indices $l$, $n$, $n'$ run over integers from $0$ to $(N-1)/2$
corresponding to the positive solutions of the BA equations.
Anderson calculated this determinant using an algorithm due to Cauchy;
his result reads~:

\begin{equation}
Z^{\pm}\propto N^{-\left(\delta_F/\pi\right)^2}
\end{equation}

\noindent
$\delta_F$ is the phaseshift at the Fermi surface which can be expressed
in terms of the density of states $n_F=1/\pi k_F$~:

\begin{equation}
\delta_F=-\arctan\left(U\pi\frac{n_F}{2}\right)
\end{equation}

\noindent
The exponent for the lattice has the same form but with the corresponding
density of states. We can now compare our result with the known result for
the static impurity. So in the $m_h=1$ case, we have~:

\begin{equation}
Z\propto N^{-\beta^+-\beta^-}~~{\rm where}~~\beta^+=\beta^-=
\left(\frac{\delta_F}{\pi}\right)^2
\end{equation}

\noindent
In contrast in the infinite mass case~:

\begin{equation}
\beta^+=\left(\frac{\delta_F}{\pi}\right)^2,~~{\rm with}~~
\delta_F=-\arctan\left(U\pi n_F
\right)~~{\rm and}~~\beta^-=0
\end{equation}

For $m_h\rightarrow\infty$, the only contribution to the
exponent comes from the even parity, the odd phaseshift vanishing.
For $m_h=1$, both parities contribute equally; moreover, the
density of possible excitations at the Fermi surface is reduced by
a factor of 2. This difference can be understood easily by simple
perturbative arguments. For an infinite mass particle, both forward
and backward scattering at the Fermi surface involve low energy excitations;
for $m_h=1$ however, backward scattering is forbidden because the
recoil energy of the particle is of the order of the
Fermi energy. So only half of the excitations are allowed.
In figure 1 we show the exponent for the two cases of equal masses
and infinite mass as a function of the dimensionless parameter
$Un_F$.

\section{Spectral function}

In this section we use our BA eigenstates to evaluate the low
energy part of the spectral function $A(Q,\omega)$; it is defined
as the Fourier Transform of the propagator of the particle
$G_Q(t)$~:

\begin{eqnarray}
A(Q,\omega)=\frac{1}{\pi}Im\int_{-\infty}^\infty e^{i\omega t}G_Q(t)
\nonumber\\
G_Q(t)=i<T\left\{b_Q(t)b_Q^{\dagger}(0)\right\}>
\end{eqnarray}

\noindent
$b_Q^{\dagger}$ ( resp. $b_Q$) is the creation (annihilation) operator
of the particle in a free state of momentum Q.
The propagator can be expanded in the eigenstates of the interacting
system~:

\begin{equation}
G_Q(t)=\sum_{\lambda}e^{-i(E_{\lambda}-\tilde E_0)t}
|\langle\tilde \psi_0(Q)|\psi_{\lambda}(Q)\rangle|^2\theta(t)
\end{equation}

\noindent
$|\tilde \psi_0(Q)\rangle$ is the non interacting ground state of N
fermions with the particle at momentum Q; its energy
is $\tilde E_0$. $|\psi_{\lambda}(Q)\rangle$ are the eigenstates of the
interacting system with energy $E_{\lambda}$ and total momentum Q.

The spectral function describes the photoemission spectrum in the
so called intrinsic approximation \cite{alm}. For an infinite mass particle
it has no momentum dependence and is totally incoherent \cite{sun}~: it has
no quasiparticle peak because of the vanishing spectral weight and
presents a power law singularity~:

\begin{equation}
A(\omega)\propto\frac{1}{\mid\omega-\omega_0\mid^{1-\beta^+}}
\end{equation}

In this section we show that the spectral function for
the $m_h=1$ has
the same low energy behaviour for momenta close to the bottom
of the band $Q\simeq0$; we treat the case $Q=0$ in detail
and mention the results for other total momenta.
We examine first how the low energy spectrum of the interacting
system with $Q=0$ can be described in the language of the non-interacting
system.

We are going now to discuss separately the case of a repulsive ($U>0$)
and attractive interaction ($U<0$); the difference is the presence of
a bound state for the attractive potential.

\subsection{$U > 0$}

There are two types of excitations : `single particle' excitations
where one of the quantum numbers $n_j$ is changed and collective
excitations where $\Lambda$ is changed. In general, we must combine
both types to have an eigenstate of zero total momentum.
We begin our discussion with the simplest excited states that
are exactly single particle like and then extend our description
to the others.

For the states with a symmetric distribution of the half integers
$n_j+1/2$ around the origin, we can achieve a zero total
momentum with $\Lambda=0$ and the distribution of momenta is
symmetric as well.
Moreover when $\Lambda=0$, the BA equations are decoupled and the
excitations are strictly single particle like, in the sense that
the spectrum corresponds to the ground state momentum distribution plus
a symmetric momentum distribution of `particle-hole' like excitations.

The situation is more complicated for the states with an asymmetric
distribution of quantum numbers because zero total momentum
implies $\Lambda\ne 0$. This causes a global
shift of all the momenta compared to the ground state and these
excitations are not strictly single particle like. However the excitation
energy and spectral weight can be approximated in order to recover
the single particle description.

Let's take the simplest type of these $\Lambda\neq 0$ excitations,
namely when one of the
$n_j$ close to the Fermi surface is changed : $n_j'-n_j=J$. If $J$ is
small compared to N the excitation energy is small and $\Lambda$ is
small as well. A similar analysis of the spectrum as in Ref. \cite{mcg}
can be performed; the main results are as follows :

\begin{eqnarray}
L(k_l'-k_l)&&= \frac{4}{U}\frac{1}{1+(2k_l/U)^2}\Lambda+\frac{16}{U^3}
\frac{k_l}{1+(2k_l/U)^2}\Lambda^2 ~~,~~ l\neq j\nonumber\\
\Lambda &&=-\frac{2\pi J}{L}\frac{\pi}{2\arctan(2k_F/U)}
\end{eqnarray}

In order to evaluate the propagator, we need to know the
excitation energy and the spectral weight. The change in energy is~:

\begin{eqnarray}
E_{\lambda}-E_0&&=k_F\frac{2\pi J}{L}
+\left(\frac{2\pi J}{L}\right)^2\frac{1}{m^*}
+O\left(\frac{1}{L^2}\right)\nonumber\\
1/m^*&&= \frac{\pi}{2}\left(\arctan(2k_F/U)-
\frac{2k_F}{U(1+(2k_F/U)^2)}\right)\Bigl/(\arctan(2k_F/U))^2
\end{eqnarray}

\noindent
The excitation energy $E_{\lambda}-E_0$ is a sum of two contributions,
the first corresponding to the particle-hole excitation and
the second to a rearrangement of the Fermi sea, a backflow term.
As pointed out by McGuire the momentum distribution
of the particle is centered around $K=2\pi J/L$ so that
we can interpret the usual backflow as a recoil of the
particle to the particle-hole excitation in the Fermi sea;
its mass is renormalized to $m^*$.
For low lying excitations $E_{\lambda}-E_0 \ll 1/2k_F^2$ so that
$\left|2\pi J/L \right| \ll k_F$. In that case the recoil energy is
negligible and the excitation energy is of particle-hole type only.

The shift of all the momenta ($k_l\rightarrow k_l'$)
influences also
the spectral weight; as for the ground state, it is a simple
determinant of the individual overlaps $X_l^p$.
Because of the particle-hole excitation $n_j\rightarrow n_j'$,
the corresponding overlap $X_j^p$ is replaced by ${X_j^p}^{\prime}$;
this is the usual effect of an
excitation in an independent particle problem. In addition, all the
other overlaps are influenced by the backflow. Using (25) we find
for $j\neq l$~:

\begin{eqnarray}
\left|\frac{X_l^p-{X_l^p}^{\prime}}{X_l^p}\right|\simeq\frac{2\pi |J|}
{Lk_F}~
\left|\frac{1}{2(n_l-p)-1}\right|&&\ll 1~~~{\rm for~} k_l/U\rightarrow 0~~
{\rm ~and~} k_l\sim k_F\nonumber\\
\left|\frac{X_l^p-{X_l^p}^{\prime}}{X_l^p}\right|\simeq\frac{2\pi|J|}
{Lk_F}&&\ll 1~~~{\rm for~ } k_l/U\rightarrow \infty
\end{eqnarray}

Thus the spectral weight has also the usual independent particle form~:
if one particle-hole pair is created around the Fermi surface, only
one of the overlaps is significantly altered.
All the low lying excitations are additive in the sense that their excitation
energy and momentum is a sum of individual contributions.
This allows us to generalize our results to all low lying excitations.

In summary we note that the backflow is not important in our problem, neither
for the excitation energy nor for the spectral weight.
This is simply due to the fact
that for a scattering between the particle at the bottom
of the band and one electron at the Fermi surface, for small momentum transfer
$q$, the recoil energy $E=q^2/2$ is negligible
compared to the particle hole energy which is linear in $q$.
We stress that this is no more
the case when the particle does not lie at the bottom of the band.

Now the propagator can be approximated in the following way :

\begin{eqnarray}
G_0(t)\simeq\ && \sum_{\left\{n_l\right\}}e^{i\sum_l
(k_l^2-\tilde k_l^2)
(t+i\xi _0^{-1})/2}\left(\det\left(X_{n_l}^{p}\right)\right)^2
\nonumber\\
= && \det\left(\sum_{l=-\infty}^{\infty}e^{i(k_l^2-\tilde k_p^2)
(t+i\xi _0^{-1})/2}X_l^p X_l^{p'}\right)
\end{eqnarray}

\noindent
where we introduced an energy cuttoff $\xi_0$ corresponding to the
range of validity of our single particle description. The
momenta $\tilde k_p$ are the non interacting one and the $k_l$ are
solutions of the BA equations for $\Lambda=0$ ie when we neglect
the backflow term.
In this last expression we can again use a basis of definite parity;
this was not possible for the exact eigenstates but our description of
the low lying excitations allows us to recover this symmetry like in
the ground state. The propagator factorizes then in two equal contributions
for odd and even parity. The calculation of this determinant was performed
by Nozi\`eres and Combescot and we only quote here their result \cite{com}:

\begin{equation}
G_0(t)\simeq ie^{-it\omega_0}
\left(i\xi_0t\right)^{-\beta^+-\beta^-}\Theta(t)
\end{equation}

\noindent
where $\beta^+=\beta^-=\left(\tilde \delta(\epsilon_F)/\pi\right)^2$
and $\tilde \delta$ is a determination of the phaseshift at the Fermi surface.
Therefore the spectral function has a divergence at an energy
$\omega_0=-(2/\pi)~\int_0^{\epsilon_F}\tilde \delta(\epsilon) d\epsilon$
and the exponent is $1-\beta^+-\beta^-$. The determination $\tilde
\delta$ is a priori
unknown, but if we compare the threshold energy $\omega_0$ with the correct
energy shift due to the interaction $E_0-\tilde E_0=-(2/\pi)~\int_0^
{\epsilon_F}\delta(\epsilon) d\epsilon$
we can fix the
determination~: $\tilde \delta=\delta$.
In summary, the spectral function at the bottom of the band
has a power law singularity and we recover the exponents
$\beta^+$ and $\beta^-$ of the orthogonality catastrophe.

What changes if the momentum is not strictly zero but still much
smaller than $k_F$~? The energies are simply shifted by a recoil term
of the particle $Q^2/2m^*$; the overlaps
$X_l^p$ around the Fermi surface are not affected as
well and the form (29) of the propagator remains~:

\begin{equation}
G_Q(t)\simeq e^{-iQ^2t/2m^*}G_0(t)
\end{equation}

The spectral function is then only shifted rigidly by the
recoil energy : its exponent is the same. Nevertheless this
is true only close to the bottom of the band : the numerical
calculations showed that this shape was strongly altered in
the vicinity of the Fermi surface \cite{dav}.

\subsection{$U < 0$}
What changes for the attractive potential ? The main difference
comes from the appearance of a bound state. Indeed the BA equations
have two imaginary solutions which in the large $L$ limit are
 \cite{mcg}~:
\begin{equation}
\kappa = \Lambda \pm i\frac{U}{2}
\end{equation}

\noindent
They always appear in pair in order to have a real total momentum.
The ground state is reached when $\Lambda =0$; the other momenta
are solutions of the BA equations with $U<0$ and the integers $n_j$ are
consecutive from $-(N-1)/2$ to $+(N-3)/2$. We would like
to note that the phaseshifts do not behave like in 3D : the appearance
of a bound state is not characterized by a phaseshift going to $\pi$
at the bottom of the band but it is simply opposite to the
phaseshift for the repulsive potential with comparable strength $|U|$.

\noindent
The ground state wave function is simple as well : it looks similar
to the repulsive case but with one of the $\phi_j$'s describing a bound
state~:

\begin{equation}
\phi_0(x)=A_0\cosh\left(\frac{U}{2}(x-\frac{L}{2})
\right),~~~~~ A_0\propto e^{-UL/2}
\end{equation}

All the analysis is similar to the repulsive case; we should
simply add one bound state in the single particle like excitation;
its overlap with the free states is simply :

\begin{equation}
X_B^m=\frac{\alpha_B}{\tilde \epsilon_m-\epsilon_B}
\end{equation}

\noindent
B stands for bound state, $\epsilon_B=\kappa^2/2$,
$\alpha_B=\cosh(\kappa L/2)\kappa/\sqrt{A_0}$ and $\tilde \epsilon_m$
is the free state energy.

Nozi\`eres and Combescot studied the influence of the
bound state and they showed that the spectral function had two
divergences, one at the energy of the true ground state, the
other at the energy of the lowest excited state which does not
contain any bound state. We first consider the absolute
threshold.

We can perform the same analysis of the low lying excitations and
end up with an asymptotic form of the propagator which is valid
for energies close to the ground state energy~:

\begin{equation}
G_0(t)\simeq e^{i\epsilon_Bt}\det\left(\sum_{l=-\infty}^
{\infty}e^{i(k_l^2-\tilde k_p^2)
(t+i\xi _0^{-1})/2} X_l^p X_l^{p'}+e^{i
(\epsilon_B- \tilde k_p^2/2)}X_B^pX_B^{p'}\right)
\end{equation}

The threshold takes place at an energy $\omega_0=2\epsilon_B-(2
/\pi)~\int_0^{\epsilon_F}\delta(\epsilon) d\epsilon$
and the exponents are the same as for the repulsive potential.

There is also a secondary threshold; we can find an asymptotic
form of the propagator taking into account all the
excited states in which the bound state is absent. The threshold
 takes place
at an energy $\omega_0=2\epsilon_F-(2/\pi)~
\int_0^{\epsilon_F}\delta(\epsilon) d\epsilon$
and the exponent is $1-\beta^+-\beta^-$ with
$\beta^+=\left(\delta_F/\pi-1\right)^2$
and $\beta^-=\left(\delta_F/\pi\right)^2$

\section{Conclusions}

In this article we calculated the exponent of the spectral function
in the asymptotic low energy range using the BA wavefunctions.
The spectral function has no
quasiparticle peak and its incoherent part has a power law divergence
at threshold. It is interesting to note that this singularity
is accompanied by an orthogonality catastrophe with similar exponents;
although such a coincidence was expected from perturbative analyses,
no exact relation between these two quantities exist.
These features are reminiscent of the
static impurity problem although the exponents are different.
This similarity however does not mean that the particle is
localized in our case; in fact the effective mass is finite
fo finite $U$ \cite{mcg,zot}. Care must be taken in
distinguishing localization as probed by transport or
optical experiments.

Recently, Ogawa {\it et al} \cite{oga} calculated the
absorption spectrum for a Tomonaga-Luttinger model; they
claimed that the exponents of the singularity did not depend
on the mass of the particle; this was also valid
for the spectral function. However our results show that
the exponents differ in the two extreme cases $m_h=\infty$
and $m_h=1$; in fact the authors neglected the
backscattering and this process becomes relevant in
the infinite mass limit as we already pointed out in section 3.
We cannot exclude however that their results might be correct
for any finite mass because the infinite mass problem
is very particular due to the broken translation symmetry. A further
analysis of the problem for intermediate masses is then needed.
As no exact solution exist for other masses, we are extending
our study of the spectral weight using Quantum Monte Carlo
techniques.

We mention the agreement of our results with the calculation in Ref.
\cite{fra} using conformal invariance~: we recover their exponents in the
presence of a strong magnetic field where the ground state of the full
Hubbard model is almost ferromagnetic.
Our results are also in good qualitative agreement with numerical
results on this model \cite{dav}; however no quantitative comparison can
be performed because of the small systems studied numerically.

Eventually, we can draw some conclusions concerning the experiments
performed on quantum wires. Although we didn't calculate the absorption
spectrum, we have noted the similarity of our situation with the
core level problem and in the latter, the divergence of the spectral
function is closely related to the edge singularities; thus the
divergence of the spectral function in our model is consistent with
the interpretation of the experimental spectra in terms of excitonic
effects.
The calculation of the absorption spectrum with the BA solutions is
under investigation.

\section{Acknowledgments}
We would particularly like to thank R. Car, F. Mila and D. Baeriswyl
for many useful discussions.
This work was supported in part by the Swiss National Science Foundation
under Grants No. 21-31144.91, No. 20-30272.90 and
the University of Fribourg.

\section{Appendix}

We prove here the assertions (11-13) which constitute central results
of this paper. Suppose that we have found a set $\left\{k_j
\right\}$ which are solutions of the B.A. equations.
We first evaluate the overlap between two plane waves built up
with different momenta $k_1$ and $k_2$.

\begin{equation}
\int_0^Le^{i((k_1-k_2)x+\delta_1-\delta_2)}dx=e^{i(\delta_1-\delta_2)}
\frac{e^{i(k_1-k_2)L}-1}{i(k_1-k_2)}=-4\frac{\sin(\delta_1)
\sin(\delta_2)}{U}
\end{equation}

\noindent
where we have used the following property of the B.A. solutions~:

\begin{equation}
\frac{1-e^{i(k_1-k_2)L}}{(1-e^{ik_1L})(1-e^{-ik_2L})}=\frac{i(k_1-k_2)}{U}
\end{equation}

\noindent
We turn now to the evaluation of the normalization factor $A_j$~:

\begin{equation}
A_j^2=\frac{1}{L}\biggl\{1+\frac{4\sin(\delta_j)^2}{LU}-2
\frac{\sin(\delta_j)}{\sum_l\sin(\delta_l)}^{-1}+
\frac{(N+1)\sin(\delta_j)^2}{\left(\sum_l \sin(\delta_l)\right)^2}
+O\left(\frac{1}{L^2}\right)\biggr\}
\end{equation}

\noindent
The sum $\sum_l\sin(\delta_l)$ is of order $L$ and we recover
the result (11). In order to gain insight in the scaling behaviour
of this quantity, we can evaluate the sum for the ground state~:

\begin{eqnarray}
\sum_l\sin(\delta_l) && =-\frac{L}{\pi}\int_0^{k_F}\sin\left(\arctan\left(
\frac{U}{2k}\right)\right)dk\nonumber\\
&&=-\frac{LU}{2\pi}\log\left(\frac{2k_F}{U}+\sqrt{\left(\frac{2k_F}
{U}\right)^2+1}\right)
\end{eqnarray}

\noindent
We note that the corrections increase with U and inversely decrease with
the density of fermions. This is generic for all the scaling behaviours
we studied in this model.

We perform now the overlap between two functions $\phi_j$ and $\phi_n$
with $j\neq n$~:

\begin{equation}
\int_0^L\phi_j^*(x)\phi_n(x)dx
=A_jA_n\biggl\{\frac{4\sin(\delta_j)
\sin(\delta_n)}{U}
-\frac{L(\sin(\delta_j)+\sin(\delta_n))}{\sum_l \sin(\delta_l)}
\biggr\}
\end{equation}

\noindent
As the normalization factors behave like $O\left(1/\sqrt{L}\,\right)$ these
overlaps decrease as well like $1/L$.

We end up the discussion with the asymptotic form of the Fourier components
of our functions $\phi_j$ which are essential in the spectral analysis~:

\begin{eqnarray}
\frac{1}{\sqrt{L}}~\int_0^L\,e^{-i\frac{2\pi m}{L}x}\phi_j(x)dx
=&&\frac{A_j \sqrt{L} \sin(\delta_j)}{\pi}\biggl(\frac{1}
{(n_j-m)-\delta_j/\pi}\nonumber\\
&&-\frac{1}{\sum_l \sin(\delta_l)}~\sum_p \frac{\sin(\delta_p)}
{(n_p-m)-\delta_p/\pi}\biggr)
\end{eqnarray}

\noindent
If $m$ is close to the Fermi surface at $n_F=(N-1)/2$, the last sum
gives a logarithmic correction~:

\begin{equation}
\sum_l\,\frac{\sin(\delta_l)}{(n_l-m)-\delta_l/\pi}\simeq\sin(
\delta_m)\log(n_F+m)
\end{equation}

The correction is then of order $\log N/N$ which decreases slowly and
tends to shift the scaling region to larger sizes in any evaluation
of the spectral properties for finite systems.

\figure{
Exponent $\beta^++\beta^-$ of the orthogonality catastrophe
as a function of the dimensionless parameter $Un_F$ where $U$
is the interaction strength and $n_F$ the density of states at the Fermi
energy for the equal masses case (solid line) and the infinite mass
case (dashed line).}
\end{document}